\documentclass[twocolumn]{aastex631}
\usepackage{amsmath}	
\usepackage{booktabs}

\shorttitle{The 2018 outburst of MAXI~J1820+070}
\shortauthors{Fan et al.}

\begin{document}

\title{The 2018 outburst of MAXI~J1820+070 as seen by {\it Insight}-HXMT}

\author{Ningyue~Fan}
\affiliation{Center for Astronomy and Astrophysics, Center for Field Theory and Particle Physics, and Department of Physics,\\
Fudan University, Shanghai 200438, China}

\author{Songyu~Li}
\affiliation{Center for Astronomy and Astrophysics, Center for Field Theory and Particle Physics, and Department of Physics,\\
Fudan University, Shanghai 200438, China}

\author{Rui~Zhan}
\affiliation{Center for Astronomy and Astrophysics, Center for Field Theory and Particle Physics, and Department of Physics,\\
Fudan University, Shanghai 200438, China}

\author{Honghui~Liu}
\affiliation{Center for Astronomy and Astrophysics, Center for Field Theory and Particle Physics, and Department of Physics,\\
Fudan University, Shanghai 200438, China}

\author{Zuobin~Zhang}
\affiliation{Center for Astronomy and Astrophysics, Center for Field Theory and Particle Physics, and Department of Physics,\\
Fudan University, Shanghai 200438, China}

\author{Cosimo~Bambi}
\affiliation{Center for Astronomy and Astrophysics, Center for Field Theory and Particle Physics, and Department of Physics,\\
Fudan University, Shanghai 200438, China}
\affiliation{School of Natural Sciences and Humanities, New Uzbekistan University, Tashkent 100007, Uzbekistan}

\author{Long~Ji}
\affiliation{School of Physics and Astronomy, Sun Yat-Sen University, Zhuhai 519082, China}

\author{Xiang~Ma}
\affiliation{Key Laboratory of Particle Astrophysics, Institute of High Energy Physics, Chinese Academy of Sciences, Beijing 100049, China}

\author{James~F.~Steiner}
\affiliation{Center for Astrophysics, Harvard \& Smithsonian, Cambridge, MA 02138, USA}

\author{Shuang-Nan~Zhang}
\affiliation{Key Laboratory of Particle Astrophysics, Institute of High Energy Physics, Chinese Academy of Sciences, Beijing 100049, China}

\author{Menglei~Zhou}
\affiliation{Institut f\"ur Astronomie und Astrophysik, Eberhard-Karls Universit\"at T\"ubingen, D-72076 T\"ubingen, Germany}

\correspondingauthor{Cosimo Bambi}
\email{bambi@fudan.edu.cn}

\begin{abstract}
We present an analysis of the whole 2018~outburst of the black hole X-ray binary MAXI~J1820+070 with {\it Insight}-HXMT data. We focus our study on the temporal evolution of the parameters of the source. We employ two different models to fit the disk's thermal spectrum: the Newtonian model {\tt diskbb} and the relativistic model {\tt nkbb}. These two models provide different pictures of the source in the soft state. With {\tt diskbb}, we find that the inner edge of the disk is close to the innermost stable circular orbit of a fast-rotating black hole and the corona changes geometry from the hard to the soft state. With {\tt nkbb}, we find that the disk is truncated in the soft state and that the coronal geometry does not change significantly during the whole outburst. However, the model with {\tt nkbb} can predict an untruncated disk around a fast-rotating black hole if we assume that the disk inclination angle is around $30^\circ$ (instead of $\sim 60^\circ$, which is the inclination angle of the jet and is usually adopted as the disk inclination angle in the literature) and we employ a high-density reflection model. In such a case, we measure a high value of the black hole spin parameter with observations in the soft state, in agreement with the high spin value found from the analysis of the reflection features and in disagreement with the low spin value found by previous continuum-fitting method measurements with the disk inclination angle set to the value of the jet inclination angle.
\end{abstract}




\section{Introduction}

Black hole X-ray binaries (BHXRBs) are binary systems in which a central black hole accretes mass from a companion star. During the accretion process, part of the gravitational energy of the accreted material is converted into electromagnetic radiation, primarily in the X-ray band. The resulting spectra typically consist of a multi-temperature blackbody component generated in the accretion disk and a power-law component produced by inverse Compton scattering of thermal photons from the disk off free electrons in the corona, which is a cloud of hot electrons near the black hole. The disk can also be illuminated by the Comptonized photons from the corona, which results in a reflection component in the observed spectra \citep[see, e.g.,][]{2021SSRv..217...65B}.

During an outburst, the spectral states of BHXRBs can be classified as hard, soft, and intermediate states \citep{Homan_Belloni_2005}. In the hard state, the spectrum is dominated by non-thermal emission caused by the Comptonization off hot electrons in the corona. In the soft state, thermal emission from the accretion disk is dominant. Different spectral states manifest in distinct regions on the hardness-intensity diagram (HID), which can be used to trace the time evolution of BHXRBs \citep{Belloni_Motta_2011}.

The knowledge of the geometry of the accretion flow is crucial for an accurate measurement of the properties of these systems. In the standard picture, the accretion disk described by the Novikov-Thorne model \citep{Novikov} is geometrically-thin and optically-thick, and the inner edge of the disk may extend to the innermost stable circular orbit (ISCO). There is wide agreement that in the soft state, for an accretion luminosity between $\sim 5$\% and $\sim 30$\% of the Eddington limit of the source, the disk extends to the ISCO \citep{2010ApJ...718L.117S,2011MNRAS.414.1183K}, but the question of whether the accretion disk is truncated at $R \textgreater R_{\mathrm{ISCO}}$ or not in the hard state remains controversial \citep{determ_spin_truncation}. 

The X-ray emission from BHXRBs usually shows variability on different time-scales. Narrow peaks in the power density spectrum (derived from the Fourier transform of the light curve) are referred to as quasi- periodic oscillations (QPOs) \citep{VanderKlis2005}. These QPOs exhibit characteristic time scales of the system's variability, making them crucial for understanding the accretion process of black holes and the evolution of the disk and coronal geometry. QPOs are primarily categorized into low-frequency QPOs (LFQPOs), with a centroid frequency in the range 0.1-30~Hz, and high-frequency QPOs (HFQPOs), with a centroid frequency in the range 40-450~Hz \citep{Belloni2010}. LFQPOs have been observed in most transient BHXRBs \citep{Motta2015}. Various theoretical models have been proposed to explain the origin of LFQPOs, including instabilities in the accretion flow \citep{Rodriguez} and geometrical effects such as precession of the inner hot flow \citep{Stella_1998} or of the jet base \citep{Sriram,MaXiang}.

MAXI~J1820+070 is a bright stellar mass black hole in an X-ray binary discovered in the X-ray band by MAXI on 11 March 2018 \citep{MAXI_Mission,Discover_maxij182070}. Using the Very Long Baseline Array and the European Very Long Baseline Interferometry Network, the distance of the source was measured to be $2.96 \pm0.3$~kpc~\citep{Atri_2020}. The inclination angle of the orbit of the binary and the mass of the central black hole were measured to be $63\pm 3 ^\circ$ \citep{Atri_2020} and $8.48^{+0.79}_{-0.72}$~$M_{\odot}$ \citep{Torres_inclination_mass}, respectively .

A number of studies have investigated the timing and spectral properties of MAXI J1820+070. LFQPOs were detected during the hard state of the outburst \citep{Buisson2019,MaXiang,You2021}. \cite{Buisson2019} observed a drastic change in the QPO frequency during the hard state, while the inner disk radius remained relatively stable. This suggests that the variability of geometric (orbital or precession) time-scales related to the inner edge of the disk may not be the primary origin of the LFQPOs in this case. Additionally, \cite{MaXiang} identified LFQPOs above $200\,\mathrm{keV}$, and observed a significant increase in the amplitude of soft LFQPO phase lag with energy above $30\,\mathrm{keV}$. These observations pose challenges to QPO models based on the instability of the accretion disk, because we cannot expect an appreciable emission of radiation above $200\,\mathrm{keV}$ from an optically thick accretion disk.

With regard to the disk geometry, different studies arrived at different conclusions on the location of the inner disk radius using different models. For instance, reflection modeling of NuSTAR data in \cite{Buisson2019} suggests that the disk was not significantly truncated in the hard state ($R_{\mathrm{in}} \approx 5.6\,R_{\mathrm{g}}$, where $R_{\mathrm{g}}=GM/c^2$ is the gravitational radius and $M$ is the black hole mass) if the black hole has low to moderate spin. In contrast, \cite{Prabhakar2022} identified a significantly truncated disk in the hard state through an analysis of the normalization factor of the \texttt{diskbb} component. Additionally, \cite{Zdziarski2022} conducted a combined analysis of \textit{Insight}-HXMT, NuSTAR, and INTEGRAL data, revealing that the inner disk radius was truncated at a value greater than 10 $R_{\mathrm{g}}$ in the hard state. From a theoretical perspective, \cite{Kawamura2022} demonstrated that a truncated disk with hot inner flow geometry could explain the timing properties of the data.

In the present paper, we fit the spectra of the 2018 outburst of MAXI J1820+070 observed by \textit{Insight}-HXMT with different models. Through the evolution of spectral parameters, we analyze the evolution of the disk and of the corona, and the relationship between some of the spectral parameters and the LFQPO frequency. 

The paper is organized as follows. Section~\ref{observation} provides a detailed account of the observations and data reduction process, while Section~\ref{fitting} presents the applied spectral models and the corresponding results. The ensuing discussion is presented in Section~\ref{discussion}, and the conclusion is in Section~\ref{conclusion}.

\section{Observation and Data Reduction}\label{observation}

The X-ray satellite \textit{Hard X-ray Modulation Telescope} (\textit{Insight}-HXMT) was launched in 2017 \citep{Zhang2020+HXMT}.  It covers a broad energy band with its three instruments (LE: 1-12~keV; ME: 8-35~keV; HE: 20-250~keV). The broad energy band and short dead time \citep{Chen2020_LE,Cao2020_ME,Liu2020_HE} of the telescope make it an ideal facility to conduct spectral-timing studies.

There are 154 archived \textit{Insight}-HXMT observations of MAXI J1820+070 which cover the entire 2018 outburst from MJD 58191 (14 March 2018) to MJD 58412 (21 October 2018).
We use the \textit{Insight}-HXMT Data Analysis Software \texttt{HXMTDAS} ver 2.04 to generate lightcurves and spectra. The background is estimated by the standalone scripts \texttt{hebkgmap}, \texttt{mebkgmap}, and \texttt{lebkgmap} \citep{Liao2020a_HEbkg,Guo2020_MEbkg,Liao2020_LEbkg}. We screen good time intervals by considering the recommended criteria, i.e., the elevation angle $>10$ deg, the geomagnetic cutoff rigidity $>8\,\mathrm{GeV}$, the pointing offset angle $<0.1$ deg, and at least $300$ s away from the South Atlantic Anomaly (SAA). 
We group the data with a minimum count of 30~photons per bin using the \texttt{grppha} tool.

\begin{figure*}
    \centering
    \includegraphics[width=0.98\linewidth]{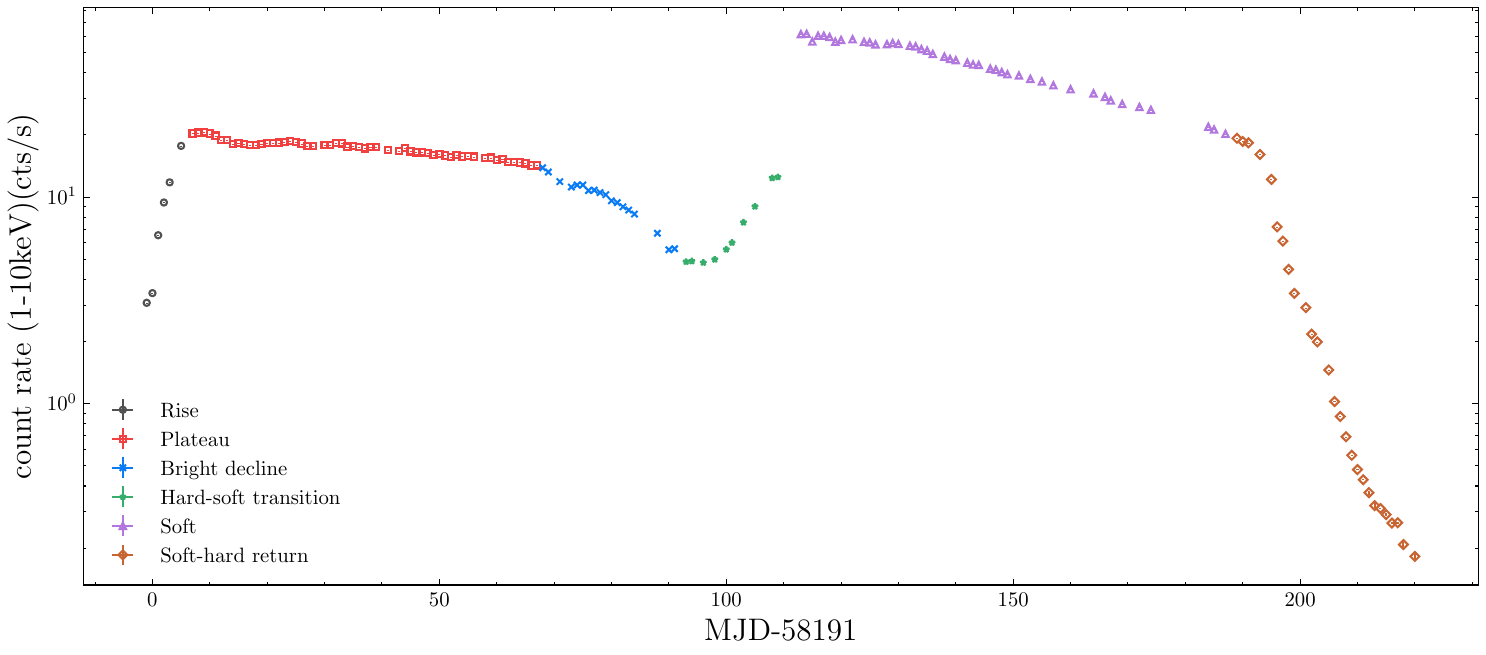}
    \caption{Light curve of the 2018 outburst of MAXI~J1820+070 detected by \textit{Insight}-HXMT. The $y$-axis shows the count rate in the 1–10~keV band (in units of counts per second). The rise, plateau, bright decline, hard-soft transition, soft, and soft-hard return phases are painted in different colors.}
    \label{lightcurve}
\end{figure*}

\begin{figure}
    \centering
    \includegraphics[width=0.98\linewidth]{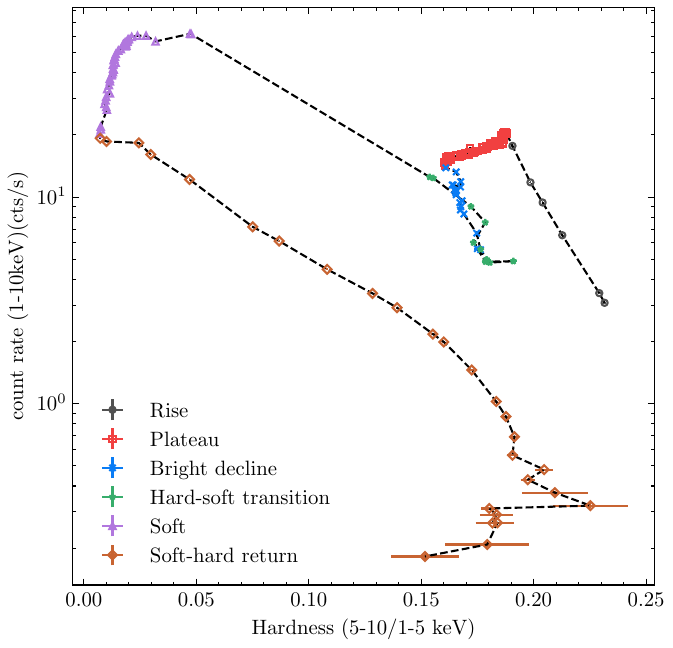}
    \caption{Hardness-intensity diagram of the whole outburst. The intensity is defined as the count rate in the 1–10~keV band (in units of counts per second) and drawn in log scale. The hardness is defined as the ratio between the count rates in the 5–10~keV band and the 1–5~keV band. Different phases are painted in different colors in accordance with Fig.~\ref{lightcurve}.}
    \label{HID}
\end{figure}

The light curve in the energy band 1-10~keV is plotted in Fig.\ref{lightcurve}, and the HID of the source is shown in Fig. \ref{HID}, from which we can clearly see different spectral states in the outburst. We divide the whole outburst into six phases: the rise, plateau, bright decline, hard-soft transition, soft, and soft-hard return phases. To analyze the evolution of the spectral parameters in different phases, spectral fitting of these observations with different models is conducted in the next section. We fit the spectra of LE in the range 1-10~keV, ME in the range 10-25~keV, and HE in the range 30-150~keV. The data in the 21-23~keV band are ignored because of calibration issues due to the silver line. We don't use the HE data in the soft and soft-hard return phases because they are dominated by background.

\section{Spectral analysis and Results}\label{fitting}

Spectral fittings are conducted with \texttt{XSPEC} v12.13.0 \citep{Xspec_Arnaud}. We implement the element aundances of \cite{abundance_Anders} and cross-sections of \cite{Verner_crosssection}. The $\chi^{2}$ statistics is used to find the best-fit values and uncertainties (at 90\% confidence level) of the parameters. A 1\% systematic error is added to the data.

\subsection{Relativistic reflection features}\label{diskbb}

We first fit the spectra with a simple continuum model: \texttt{constant $\times$ tbabs $\times$ (simplcutx $\times$ diskbb)} (Model 1). We employ the \texttt{diskbb} model \citep{Mitsuda} to fit the multi-temperature blackbody component from the accretion disk, and the \texttt{simplcutx} model \citep{Steiner2017} to fit the Comptonizated spectrum generated by the inverse Compton scattering of thermal photons from the disk off free electrons in the corona. \texttt{constant} acts as cross normalization among the LE, ME, and HE data. Additionally, we incorporate the \texttt{tbabs} model to account for interstellar absorption \citep{Wilms_tbabs}, with the column density fixed at $N_{\mathrm{H}}=0.15\times10^{22}$~cm$^{-2}$ \citep{Uttley}.

\begin{figure}
    \centering
    \includegraphics[width=0.99\linewidth]{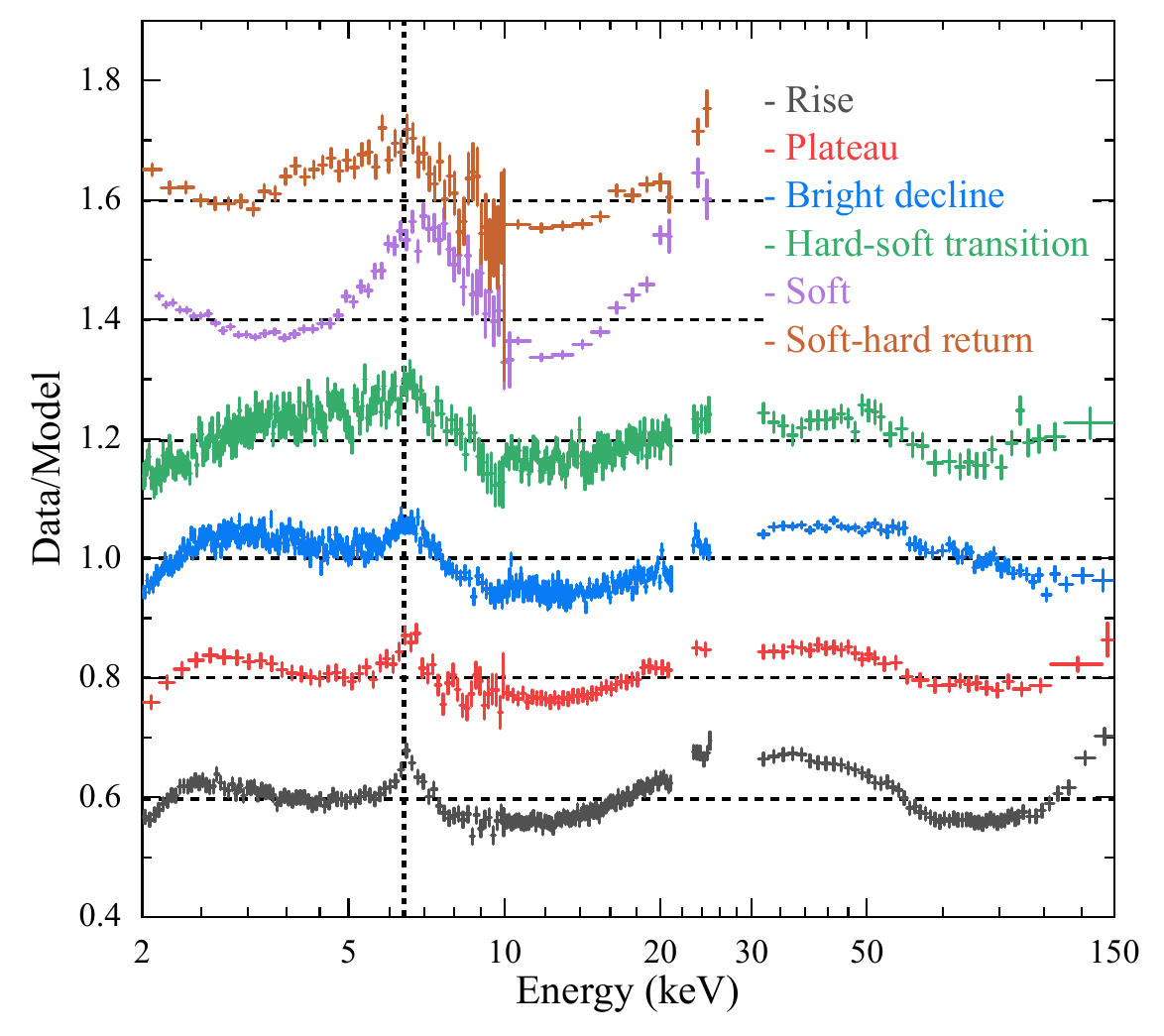}
    \caption{A selected sample of data-to-model ratios from different phases of the outburst, painted in different colors in accordance with Fig.~\ref{lightcurve}. Specifically, we show the data-to-model ratios of the observation on March 20 for the rise phase, of the observation on May 1 for the plateau phase, of the observation on June 2 for the bright decline phase, of the observation on July 1 for the hard-soft transition phase, of the observation on July 7 for the soft phase, and of the observation on October 3 for the soft-hard return phase. The model used here is \texttt{constant $\times$ tbabs $\times$ (simplcutx $\times$ diskbb)}. To make the plot clearer, the curves of different phases are parallel moved higher or lower. The data-to-model ratios show two clear peaks around 5-8~keV (iron line) and 20-30~keV (Compton hump). The vertical dotted line marks the position of neutral K$\alpha$ iron line at 6.4~keV. The data in the 21-23~keV band are ignored because of calibration issues related to the silver line. The HE data in the soft and soft-hard return phases are not used because they are dominated by background due to their low count rate.}
    \label{ironline}
\end{figure}

The ratios between data and the best-fit models of seven selected spectra from different phases are plotted in Fig. \ref{ironline}, from which we can see clear reflection features: a broadened iron emission line around 6.4~keV and a Compton hump with peak around 30~keV \citep[caused by electron down scattering of high energy photons and photoelectric absorption on the disk,][]{Fabian_Compton}. Consequently, we introduce the relativistic reflection model \texttt{relxillCp} \citep{Garcia_relxill}. The non-relativistic reflection model \texttt{xillverCp} \citep{Garcia_xillver} is also included because, in general, spectra of Galactic black holes may require even a distant reflector \citep{xillver_chakraborty}.

\subsection{Spectral fitting with a reflection model}
The model used in this part is: \texttt{constant $\times$ tbabs $\times$ (simplcutx $\times$ (diskbb + relxillCp) + xillverCp)} (Model 2). The parameters in the reflection model include the inclination angle  of the accretion disk with respect to our line of sight ($i$), the spin of the black hole ($a_{*}$), the inner disk radius ($R_{\mathrm{in}}$), the reflection fraction, the emissivity index ($q$) (we model the emissivity of the reflection component with a power law, so $\epsilon \propto r^{-q}$), the photon index of the power law component from the corona ($\Gamma$), the electron temperature in the corona ($kT_\mathrm{e}$), the iron abundance ($A_{\mathrm{Fe}}$), and the ionization parameter ($\log\xi$) of the disk. We fix the inclination angle of the accretion disk at $i=63^{\circ}$ \citep{Atri_2020} and the spin of the black hole at $a_{*}=0.998$ because there's strong degeneracy between the inner radius and the spin parameter \citep{Buisson2019} and we focus on $R_{\mathrm{in}}$ here. The reflection fraction is fixed at $-1$ so that only the reflection component is considered because the Comptonized spectrum is already described by \texttt{simplcutx $\times$ diskbb}. We link $kT_\mathrm{e}$ and $\mathrm{\Gamma}$ in \texttt{relxillCp} and \texttt{xillverCp} to $kT_\mathrm{e}$ and $\mathrm{\Gamma}$ in \texttt{simplcutx}. Since the temperature of the inner part of the accretion disk can be high in BHXRBs, non-thermal effects may not be negligible and produce deviations from a blackbody spectrum. This is normally taken into account by introducing a color factor (or hardening factor) \citep{color_correction}. However, there is no color factor parameter in \texttt{diskbb}.

\begin{figure*}
    \centering
    \includegraphics[width=0.32\linewidth]{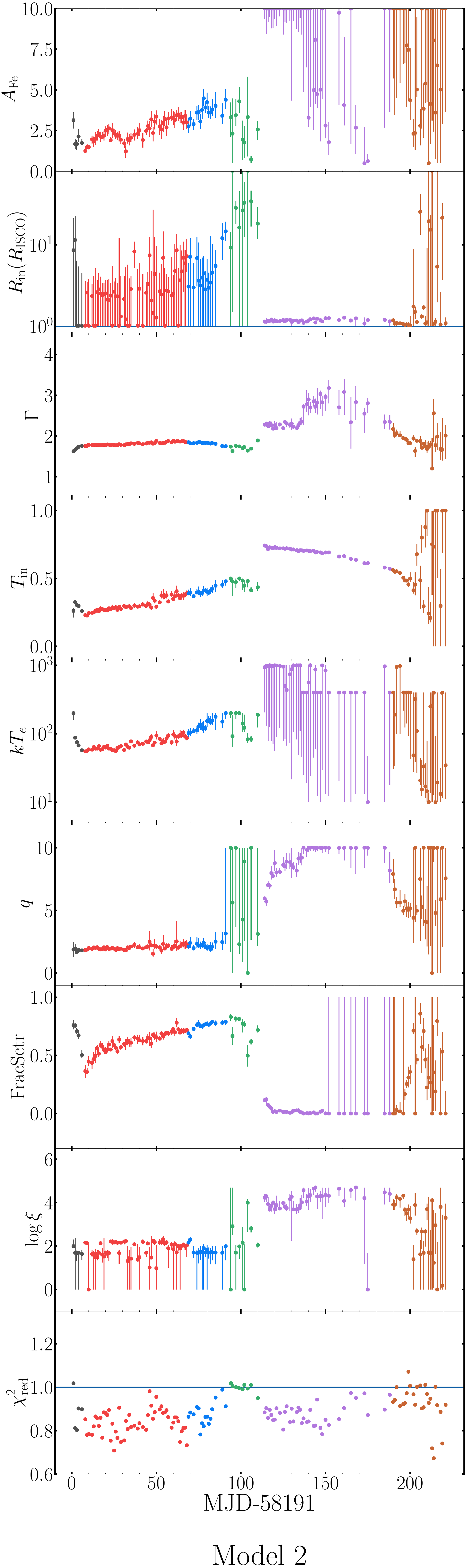}
    \includegraphics[width=0.32\linewidth]{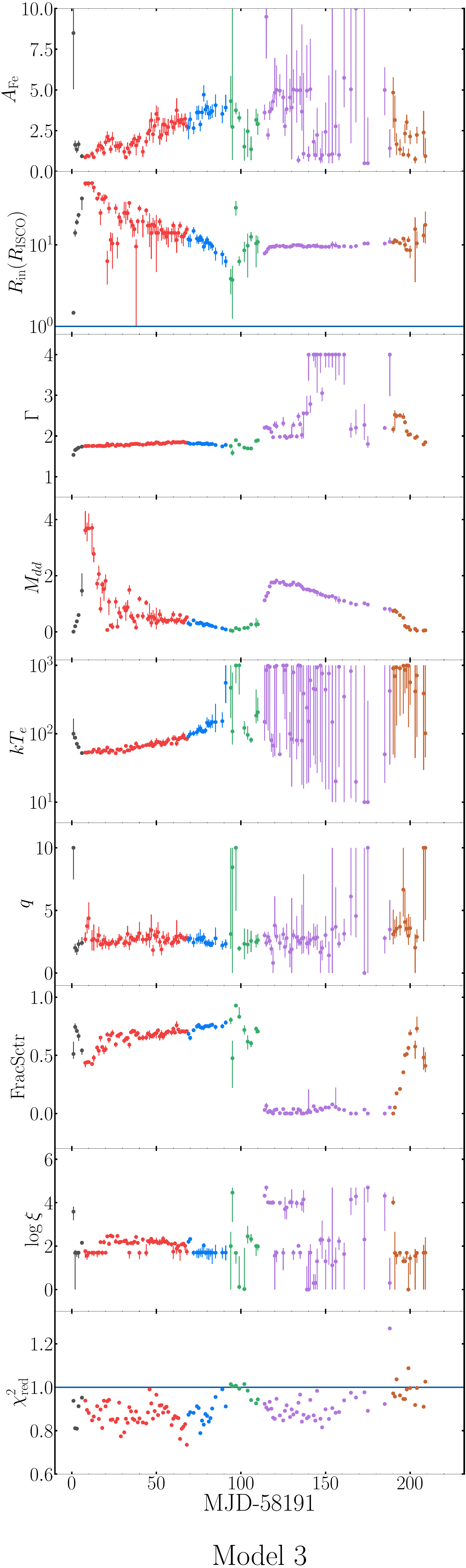}
    \includegraphics[width=0.32\linewidth]{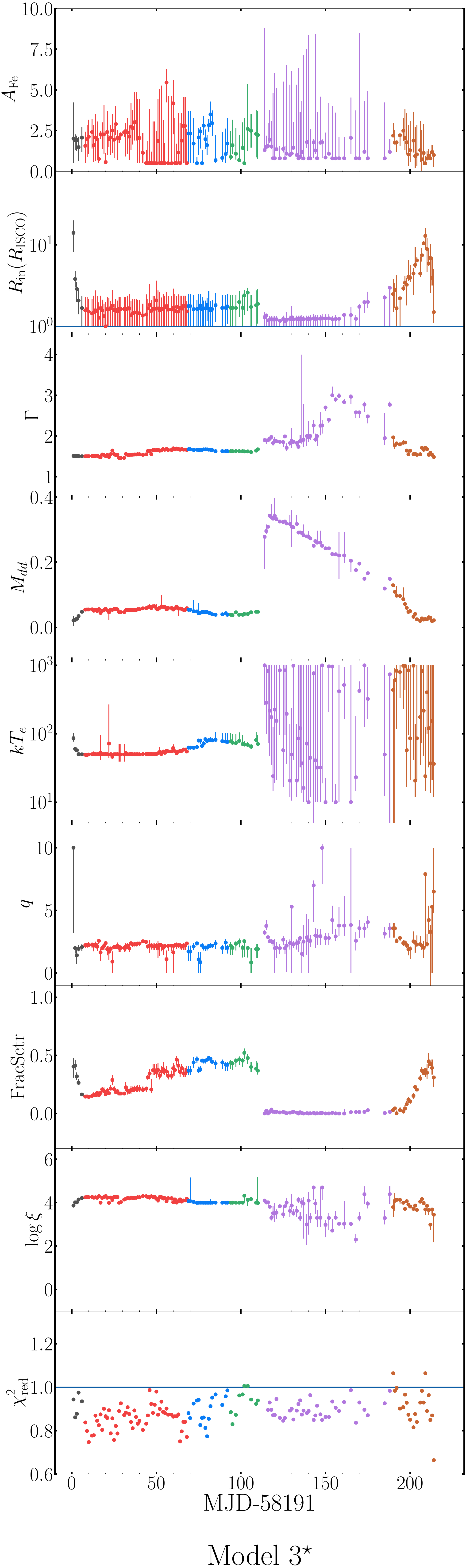}
    \caption{Temporal evolution of the parameters in Model~2 (left panel), Model~3 (central panel), and Model~3$^\star$ (right panel). Model~2 is \texttt{constant $\times$ tbabs $\times$ (simplcutx $\times$ (diskbb + relxillCp) + xillverCp)}. Model~3 and Model~3$^\star$ are both \texttt{constant $\times$ tbabs $\times$ (simplcutx $\times$ (nkbb + relxillCp) + xillverCp)} and the difference is that the inclination angle, the disk electron density, and the hardening factor are, respectively, $63^{\circ}$, $10^{15}$~cm$^{-3}$, and 1.7 in Model~3 and $32^{\circ}$, the $10^{20}$~cm$^{-3}$, 1.6 in Model~3$^\star$. The parameters are (from top to bottom): the iron abundance, the inner disk radius, the photon index, the inner disk temperature (Model~2) or the mass accretion rate (Models~3 and 3$^\star$), the electron temperature of the corona, the emissivity index, the scattering fraction, the ionization parameter, and the reduced $\chi^2$ of the best-fits.}
    \label{superfigure}
\end{figure*}

We find that most spectra can be fit well with reduced $\chi^{2}$ close to 1. The evolution of the spectral parameters is shown in the left panel of Fig.~\ref{superfigure}. The iron abundance of the accretion disk turns out to be about 3-4~times the  solar iron abundance in the hard state and surges to the upper bound in the soft state. The constraint on the inner disk radius is weak in the hard state, as was already found in \cite{Zdziarski2022}. Subsequently, the inner disk radius seems to be close to the ISCO in the soft phase, and it is again difficult to constrain during the soft-hard return phase. The inner disk temperature experiences a slight decrease during the rise phase, followed by gradual growth during the plateau and subsequent decline phases, reaching approximately 0.7~keV in the soft state. It gradually decreases as the system transitions back to the hard state. Similarly, the electron temperature of the corona exhibits a minor decrease during the rise phase, followed by a mild increase during the plateau phase. The constraint on the corona electron temperature is weak in the soft state. Regarding the photon index, it transitions from a low value ($< 2$) in the hard state to a higher one ($3-4$) in the soft state, subsequently decreasing as the system returns to a higher hardness level. The ionization parameter evolves from below 2 in the hard state to approximately 4 in the soft state, declining during the soft-hard return phase.

\subsection{Relativistic thermal component}\label{nkbb}

Since \texttt{diskbb} is a non-relativistic model, we replace it with the relativistic thermal model \texttt{nkbb} to see whether it improves the fit and provides a different estimate of some parameters. \texttt{nkbb} \citep{Menglei} is a model for relativistic thermal spectra. We use \texttt{nkbb} because (unlike \texttt{kerrbb}) the radial coordinate of the inner edge of the disk is not fixed at the ISCO radius and is a free parameter to fit. We connect this parameter with $R_{\mathrm{in}}$ of the reflection component \texttt{relxillCp} and we still keep the black hole spin parameter frozen to 0.998.

In the new model \texttt{constant $\times$ tbabs $\times$ (simplcutx $\times$ (nkbb + relxillCp) + xillverCp)} (Model 3), we set the mass of the black hole to 8.48~$M_{\odot}$ \citep{Torres_inclination_mass} and the distance to 2.96~kpc~\citep{Atri_2020}. The hardening factor of \texttt{nkbb} is fixed at 1.7.

The fitting results of the spectral parameters are shown in the central panel of Fig.~\ref{superfigure}. The evolutions of the photon index, electron temperature of the corona, and scattering fraction are consistent with the fitting results of Model 2. In the hard state, the inner disk radius is better constrained with the new model. It initially increases in the rising phase, then gradually decreases in the plateau and subsequent bright decrease phases, before increasing again in the hard-soft transition phase. In the soft phase and soft-hard return phase, it stabilizes around $10\,R_{\mathrm{g}}$, in disagreement with the fits of Model~2. 
In Model~3, the inclination angle, mass, and distance of the black hole are fixed, so the uncertainties on the best-fit values of our fits do not include the uncertainties from the measurements of these three parameters. If we set the black hole mass to 5~$M_{\odot}$ or 15~$M_{\odot}$, we obtain very similar results, so the uncertainty on the black hole mass has only a weak impact on the estimate of $R_{\mathrm{in}}$. To evaluate the impact of the distance and of the disk inclination angle on the estimate of $R_{\mathrm{in}}$, we can use the \texttt{diskbb} normalization factor $norm_{\mathrm{diskbb}}=(R_{\mathrm{in}}/D_{10})^2\cos{i}$, where $D_{10}$ is the distance to the source in units of 10~kpc. From this formula, the uncertainty in $R_{\mathrm{in}}$ caused by the uncertainties in the distance and inclination angle should be not more than about 20\%, so they cannot explain the discrepancy in the estimates of $R_{\mathrm{in}}$ between Models~2 and 3.

The emissivity index seems to remain between 2 and 3 during the whole outburst, while we had found a very high emissivity index in the soft phase in Model~2. The ionization parameter fluctuates significantly in the soft state, maintaining with a value near 2 in the hard state and the hard-soft return phase. The mass accretion rate surges in the rising phase and then sharply declines in the plateau phase. Subsequently, it remains relatively low in the remaining hard state, before increasing to approximately $2\times10^{18}\,\mathrm{g/s}$ in the soft state.

\begin{figure*}
    \centering
        \includegraphics[width=0.32\textwidth]{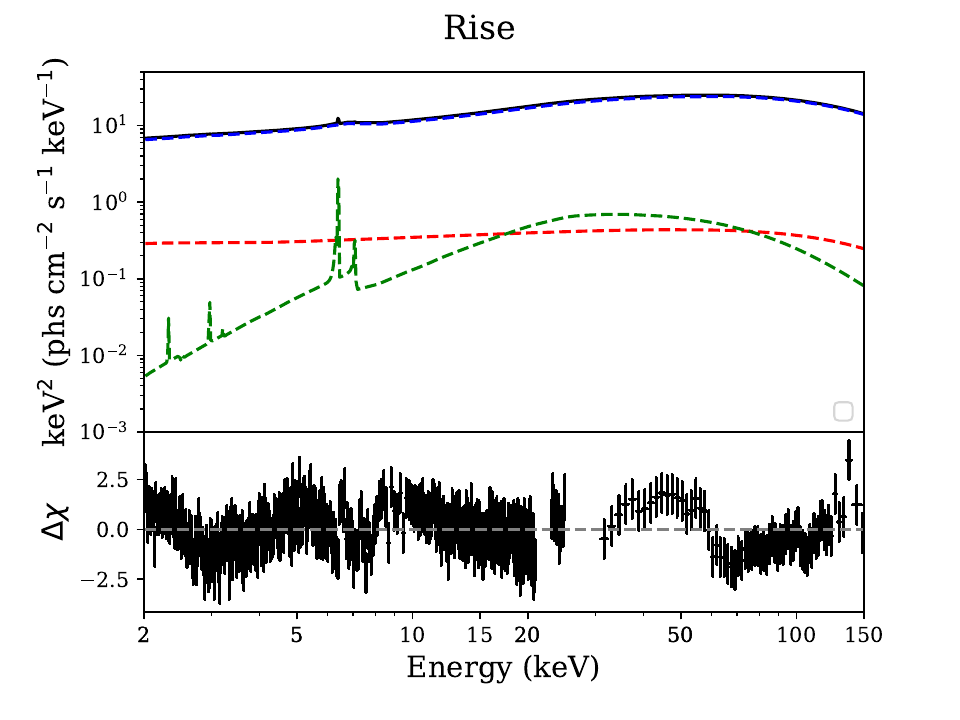}
        \includegraphics[width=0.32\textwidth]{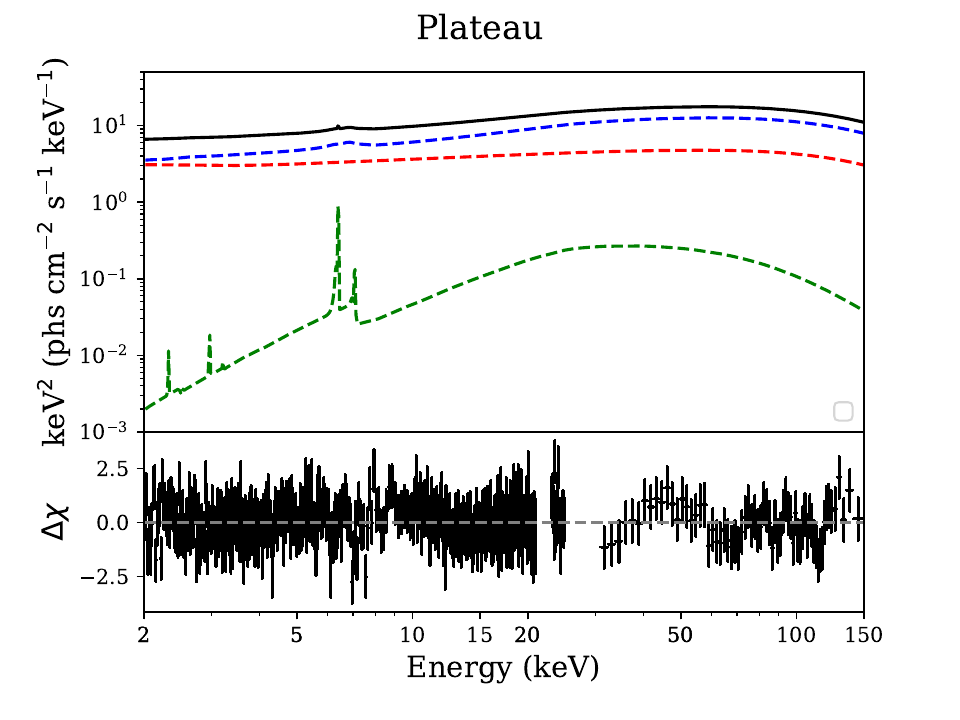}
        \includegraphics[width=0.32\textwidth]{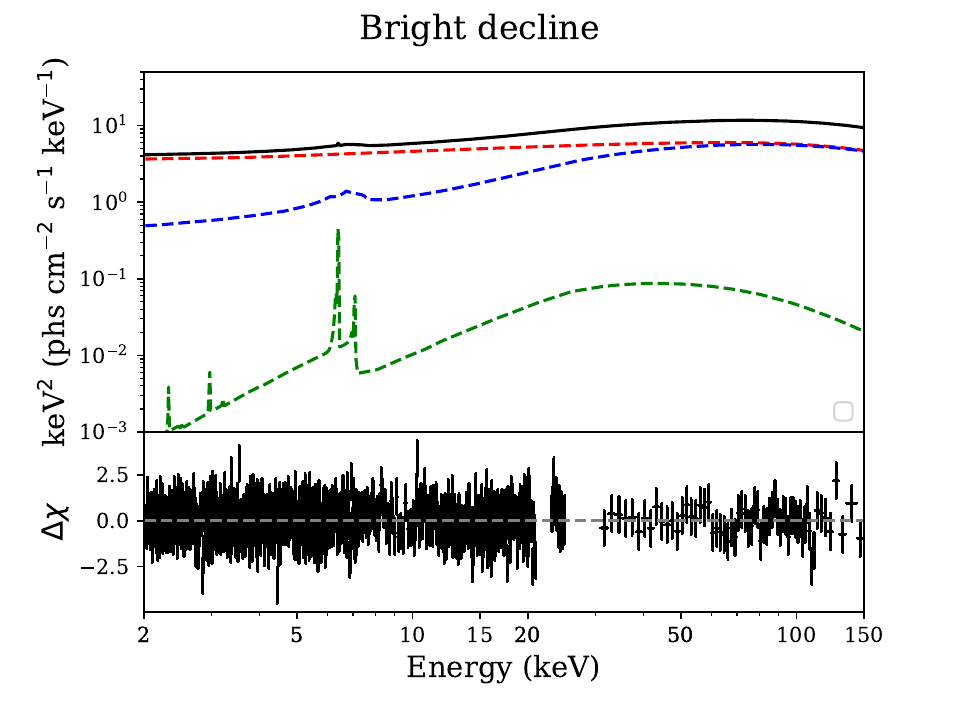}
        \includegraphics[width=0.32\textwidth]{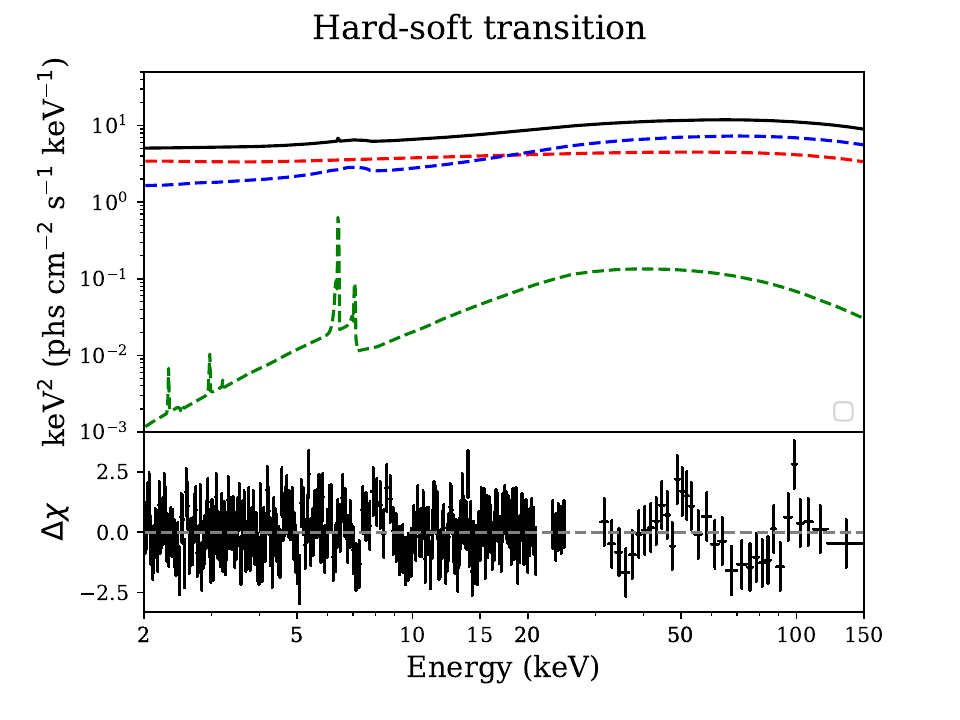}
        \includegraphics[width=0.32\textwidth]{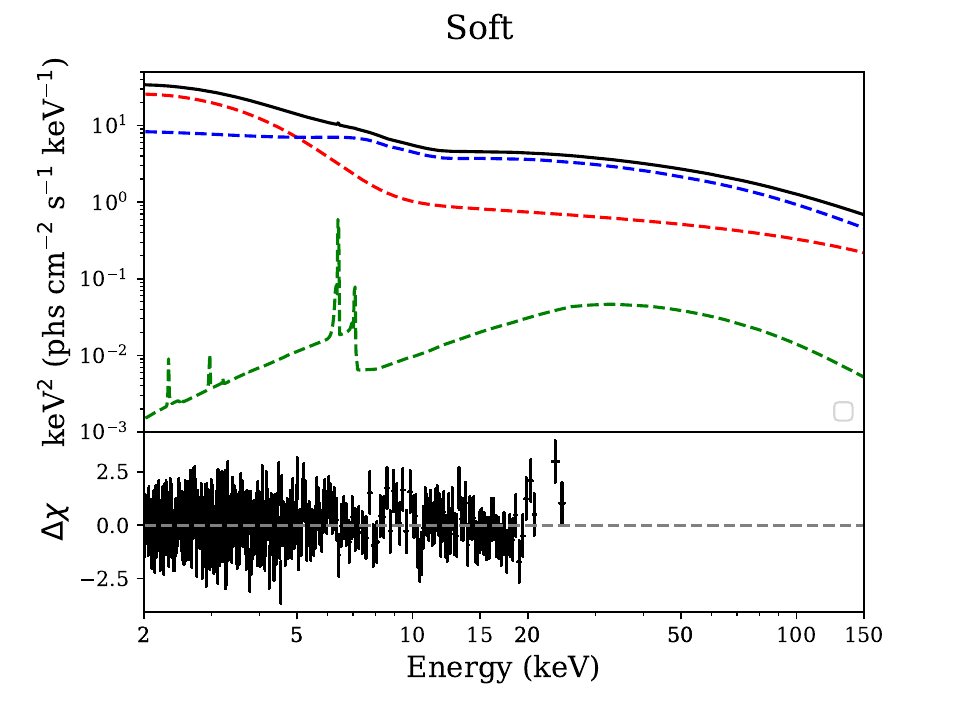}
        \includegraphics[width=0.32\textwidth]{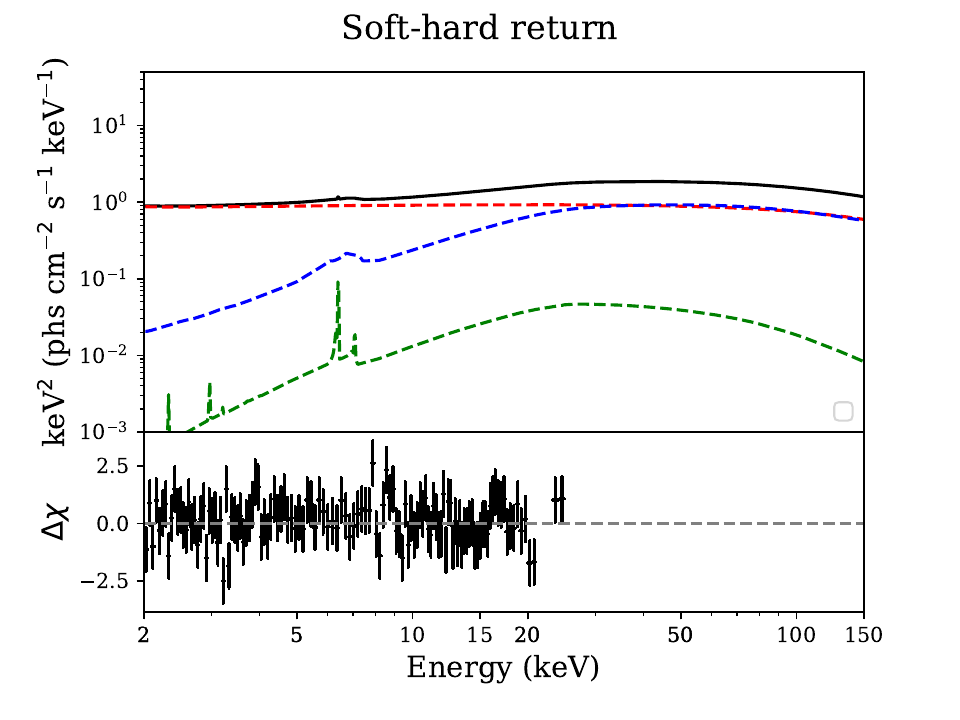}
    \caption{Spectral components and residuals of the fittings of the days picked from different phases and already shown in Fig.~\ref{ironline}; i.e., of the observations on March 20 (rise phase), May 1 (plateau phase), June 2 (bright decline phase), July 1 (hard-soft transition phase), July 7 (soft phase), and October 3 (soft-hard return phase). The model used here is \texttt{constant $\times$ tbabs $\times$ (simplcutx $\times$ (nkbb + relxillCp) + xillverCp)} (Model~3). The black solid line is for the total model. The red, blue, and green dashed lines represent the \texttt{nkbb} component, the \texttt{relxillCp} component, and the \texttt{xillverCp} component, respectively. The \texttt{nkbb} and \texttt{relxillCp} components have been scattered by \texttt{simplcutx} here. The data in the 21-23~keV band are ignored because of calibration issues related to the silver line. The HE data in the soft and soft-hard return phases are not used because they are dominated by background due to their low count rate.}
    \label{model}
\end{figure*}

We plot the spectral components and residuals of the best-fit model of days picked from different phases in Fig.~\ref{model}. In the rising phase, the reflection component is dominant, while the disk component becomes increasingly important in latter phases. In the soft phase and soft-hard return phase, the thermal disk component is the dominant component in the soft X-ray band.

\begin{table*}
\centering
    \begin{tabular}{lccc}
         & Model 2 & Model 3 & Model 3$^\star$ \\
        \hline
        Thermal model & \texttt{diskbb} \citep{Mitsuda} & \texttt{nkbb} \citep{Menglei} & \texttt{nkbb} \citep{Menglei} \\
        \hline
        Hardening factor$^1$ & N/A & 1.7 & 1.6 \\
        \hline
        Inclination angle$^{1,2}$ [deg] & 63 \citep{Atri_2020} & 63 \citep{Atri_2020} & 32 \citep{low_inclination} \\    
        \hline
        Electron density$^1$ [cm$^{-3}$] & $10^{15}$ & $10^{15}$ & $10^{20}$ \\ 
        \hline          
        Spin$^{3,4}$ & $\sim 0.98$ & $\sim 0.3$ & $\sim 0.94$ \\
        \hline
        Inner disk radius$^{4}$ & $\sim R_{\rm ISCO}$ & $\sim 10~R_{\rm ISCO}$ & $\sim R_{\rm ISCO}$ \\
        \hline
        Emissivity index$^5$ & High & $2 \sim 3$ & $2 \sim 3$ \\
        \hline
        Scattering fraction$^5$ & Very low & Very low & Very low \\
        \hline
    \end{tabular}
    \caption{Summary of the assumptions and measurements in the soft phase with different models. Notes: \\
    $^1$ These parameters are frozen in the fits.\\
    $^{2}$ There is no inclination angle in \texttt{diskbb} and the value in the table for Model~2 refers to the value of the inclination angle of the reflection components. In Model~3 and Model~3$^\star$, the values of the inclination angle in the thermal and reflection components are linked together. \\
    $^{3}$ \cite{Draghis_spin} find $a_{*}=0.988^{+0.006}_{-0.028}$ from the analysis of the reflection features with {\tt relxill}. \cite{Zhao_spin} and \cite{Guan_spin} find, respectively, $a_{*}=0.14\pm0.09$ and $a_{*}=0.2^{+0.2}_{-0.3}$ from the analysis of the thermal component with \texttt{kerrbb2} assuming the disk inclination angle $i = 63^\circ$. \\
    $^{4}$ To measure the spin, we set the inner disk radius to $R_{\rm ISCO}$ and we leave $a_*$ free. To measure the inner disk radius, we set the spin parameter to 0.998 and we leave $R_{\rm in}$ free. \\
    $^5$ In Model~2, the emissivity index is around 2-3 before the soft phase and increases dramatically in the soft phase, suggesting that the corona becomes compact and close to the black hole in the soft phase. In Model~3 and Model~3$^\star$, the emissivity index is around 2-3 over the whole outburst, suggesting that the coronal geometry does not change much over the whole outburst and the decrease of the scattering fraction in the soft phase may be attributed to the decrease of the coronal density.}
    \label{model_comparison}
\end{table*}

\section{Discussion}\label{discussion}

This section is dedicated to the discussion of our results. A summary of the main findings from the analysis of the spectra in the soft phase is reported in Table~\ref{model_comparison}.

\subsection{High iron abundance}\label{ironabundance}
 
All the above fitting results indicate a high and changing iron abundance after the system enters the soft state. However, the real iron abundance should not change during an outburst. It is still unclear how to interpret current measurements of the iron abundance from the analysis of the reflection features. For many sources, including Galactic black holes and active galactic nuclei (AGN), we find unexpectedly high iron abundances. However, there are also some Galactic black holes with low iron abundance  \cite[see, for instance, Fig. 3 in][]{Garcia_highiron}. 

For MAXI~J1820+070, a high iron abundance has also been observed in other studies \citep{Buisson2019,Zdziarski2021}. Some studies found a lower iron abundance by employing a different corona-disk model (see Fig. 5 in \cite{Kawamura2022} and Fig. 4 in \cite{Zdziarski2021}, for example). These models incorporate two reflection components: one describes a harder incident component from the Comptonization of the hot accretion flow interior to the disk reflected by the outer flared disk, and another describes a softer incident component from the hot Comptonizing corona covering the inner part of an extended disk reflected by the disk beneath it. With this model, the iron abundance can be constrained to be close to the solar iron abundance in the soft state.

To figure out the impact of the iron abundance on our results, we refit some spectra from different phases with Model~3 fixing the iron abundance to 2.5. In the case of the spectra in the soft and soft-hard return phases, we do not find significant differences in the values of $\chi^{2}$ as well as in the estimates of the parameters of the model. This is because the reflection features are weak and, even when the iron abundance is a free parameter, its value cannot be constrained well. In the other phases, we find that the quality of the fits is worse ($\chi^{2}$s are higher), but still we do not see any significant difference in the estimate of the parameters of the model.

We note that even the disk electron density influences the measurement of the iron abundance. \cite{Jiachen_iron_abundance} found that high density disk reflection models can significantly reduce the inferred iron abundances. It is also possible that some real physical mechanisms are making the iron abundance higher than expected. \cite{Reynolds_iron_abundance} introduced the hypothesis that the radiative levitation of iron ions in the inner part of the accretion disks can enhance the photospheric abundance of iron.

\subsection{Disk and corona evolution in different models}\label{geometry}

\subsubsection{Inner edge of the disk}

The fitting with Model 3 shows that the disk is highly truncated in the hard state, which is consistent with \cite{Zdziarski2022,Prabhakar2022}. In the soft state, while the fitting with Model 2 (where \texttt{diskbb} is the thermal component) indicates that the inner disk radius is near the ISCO, the fitting with Model 3 (where \texttt{nkbb} is the thermal component) reveals that the inner disk radius is about 10 $R_{\mathrm{ISCO}}$ (see Fig.~\ref{superfigure}).

In an attempt to find a consistency between these two models, we first try to replace \texttt{relxillCp} with \texttt{relxillNS} in the soft state. The incident spectrum of \texttt{relxillNS} is a blackbody component of temperature $kT\mathrm{_{bb}}$, which may describe the returning radiation of the thermal spectrum of the disk and generate the reflection spectrum in the soft state (when the corona is probably small). However, we find that the fitting results are generally very similar to the previous results with \texttt{relxillCp}. We also use the element abundances of \cite{Wilms_tbabs} and the result is still consistent. Moreover, we try to link the inner disk radius in \texttt{relxillCp} and in the \texttt{diskbb} normalization factor using the formula $norm_{\mathrm{diskbb}}=(R_{\mathrm{in}}/D_{10})^2\cos{i}$ in Model 2, or thaw the link between $R_{\mathrm{in}}$ in \texttt{nkbb} and \texttt{relxillCp} in Model 3, but $R_{\mathrm{in}}$ is still close to the ISCO in the soft state in Model 2 and is about 10 $R_{\mathrm{ISCO}}$ in Model 3. We thus conclude that the choice to link or not $R_{\mathrm{in}}$ between different components in different models cannot be the main cause of the difference in measuring $R_{\mathrm{in}}$. 

We also check if the results change by changing the values of some frozen  parameters. A different mass of the black hole ($10M_{\odot}$ and $15M_{\odot}$) doesn't have great influence on $R_{\mathrm{in}}$. However, we find that if the inclination angle is set to be around $32^{\circ}$ \citep[which is the disk inclination measured in][]{low_inclination}, the hardening factor is 1.6 and the disk electron density is $10^{20}\,\mathrm{cm^{-3}}$, the inner disk radius will be close to the ISCO in the soft state, which makes the results given by Model 2 and Model 3 consistent. The fitting of data in the whole outburst under this parameter setting is in the right panel of Fig.~\ref{superfigure} (Model~3$^\star$). We can see that in the soft state, the inner disk radius is close to the ISCO at the beginning, and grows slightly to about $2\,R_{\mathrm{ISCO}}$ at the end of the soft state.

\subsubsection{Mass accretion rate}

\begin{figure*}
    \centering
    \includegraphics[width=0.5\textwidth]{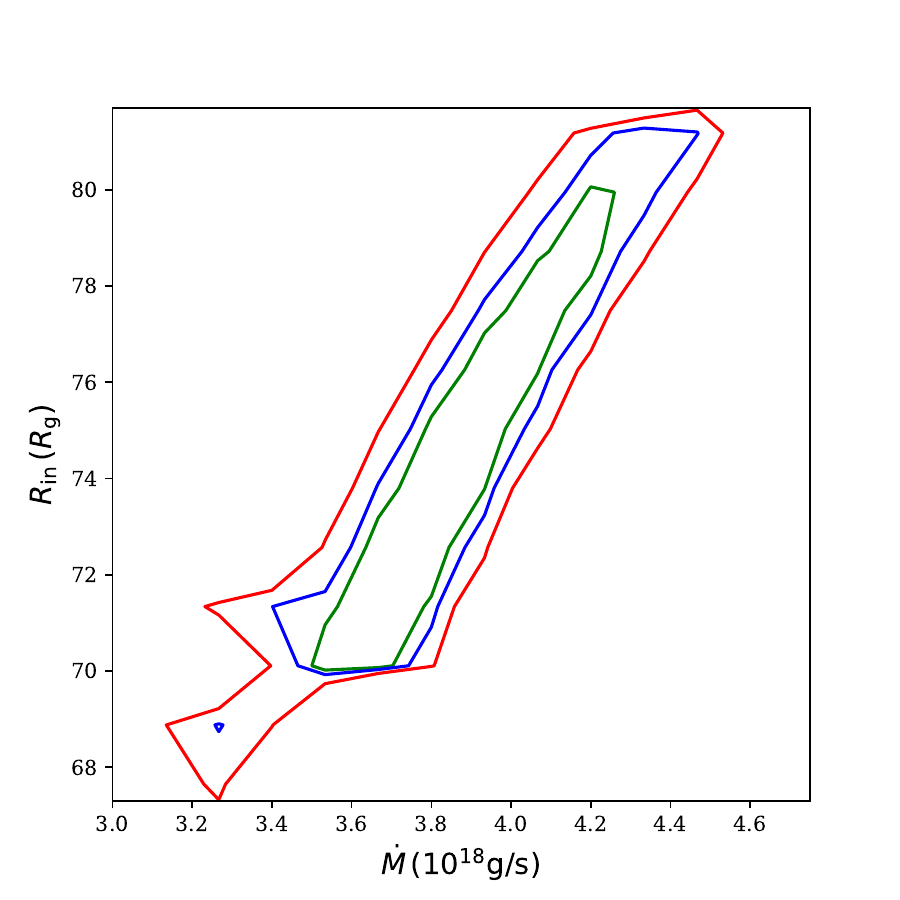}
    \caption{Correlation between the mass accretion rate and inner disk radius. The red, green, and blue lines are, respectively, the 68\%, 90\%, and 99\% confidence level contours for two relevant parameters (corresponding to $\Delta\chi^2 = 2.30$, 4.61, and 9.21, respectively).}
    \label{degeneracy}
\end{figure*}

In the fittings with Model 3, we observe a surge in the mass accretion rate during the rise phase followed by a slow decay. The inner disk radius also increases a little at first and then decreases slowly (central panel in Fig. \ref{superfigure}). Since the classical disk accretion model cannot explain the simultaneous increase of the inner disk radius and the mass accretion rate, we plot the correlation between the two parameters of a chosen day in the rising phase in Fig.~\ref{degeneracy}. There is indeed strong degeneracy between the inner radius and the mass accretion rate, which is expected since a smaller inner radius can somehow compensate the effect on the spectral shape by a lower accretion rate \citep{Honghui_QPO}. With Model~3$^\star$, the accretion process is consistent with the classical scenario: at the beginning of the outburst, the inner disk radius decreases from about 10 $R_{\mathrm{ISCO}}$ and the mass accretion rate increases a little. In the soft state, the mass accretion rate surges and then slowly decreases. At the end of the outburst, the accretion process becomes weaker again and the disk is truncated (see the right panel of Fig.~\ref{superfigure}).

\subsubsection{Evolution of the corona}

We can also infer the evolution of the corona from the spectral parameters. We model the emissivity profile of the disk with a power-law, $\epsilon \propto r^{-q}$ \citep{Peng_2023,Kara2019}. The left panel in Fig.~\ref{superfigure} shows an increase in the emissivity index $q$ when the source transitions to the soft state, suggesting that the corona becomes compact and close to the black hole. The  scattering fraction decreases in the rise phase, moderately increases in the plateau and bright decline phases, and drops to a very low value between the hard-soft transition and the beginning of the soft phase. If we combine the behavior of the emissivity index and of the scattering fraction, within Model~2 it seems like that in the hard state the corona is extended and/or not too close to the black hole, while it is compact and close to the black hole in the soft phase.

However, in Models~3 and 3$^\star$, we find a different picture, as the value of $q$ is about 2-3 during the whole outburst. This would suggest that the coronal geometry does not change and the dramatic change of the scattering fraction from the hard to soft states may be determined by the coronal density. As the system transitions back to the hard state, the scattering fraction increases in all models. We note that from the study of the evolution of coverage fraction, \cite{Peng_2023} proposed a scenario where the corona first contracts, then it is replaced by a precessing jet in the plateau phase, and eventually becomes very small in the soft state.

Some  studies on this source have employed a lamppost corona. In such a case, the corona would be a point-like source along the black hole spin axis \citep{Dauser2013_lamppost}. A temporal variation of the height of the corona may also explain the evolution of spectral parameters. \cite{Prabhakar2022} found that while the height of the corona keeps decreasing before entering the soft state, its reflection fraction first increases then decreases, indicating that the corona is possibly moving back.

\subsubsection{Other considerations}

From the fittings of different models, we see some differences in the estimates of the iron abundance $A_{\mathrm{Fe}}$,  photon index $\Gamma$ and ionization parameter $\log \xi$ in the soft state. It is likely because the coronal spectrum and the reflection component are weak in the soft state, so changing the continuum model for the disk thermal emission can impact the parameter measurements of the reflection component. The iron abundance is lower in Model~3$^\star$ than in Model~3, which is a consequence of the higher disk electron density in Model~3$^\star$ (see the discussion on the iron abundance in Section~\ref{ironabundance}).

\subsection{LFQPO characteristics}

\begin{figure*}
    \centering
        \includegraphics[width=0.32\textwidth]{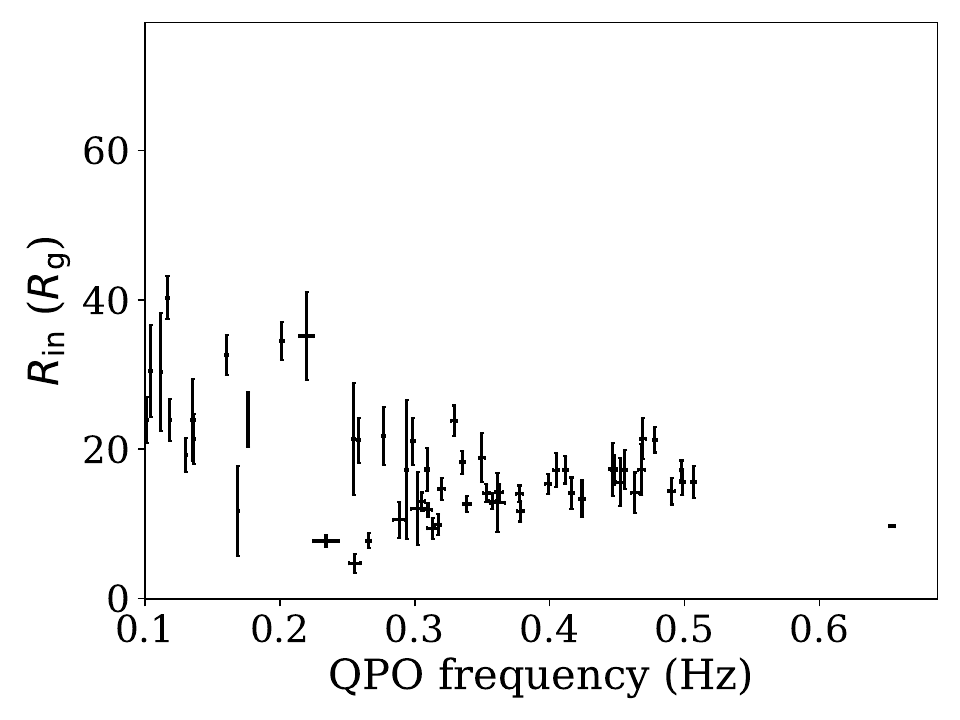}
        \includegraphics[width=0.32\textwidth]{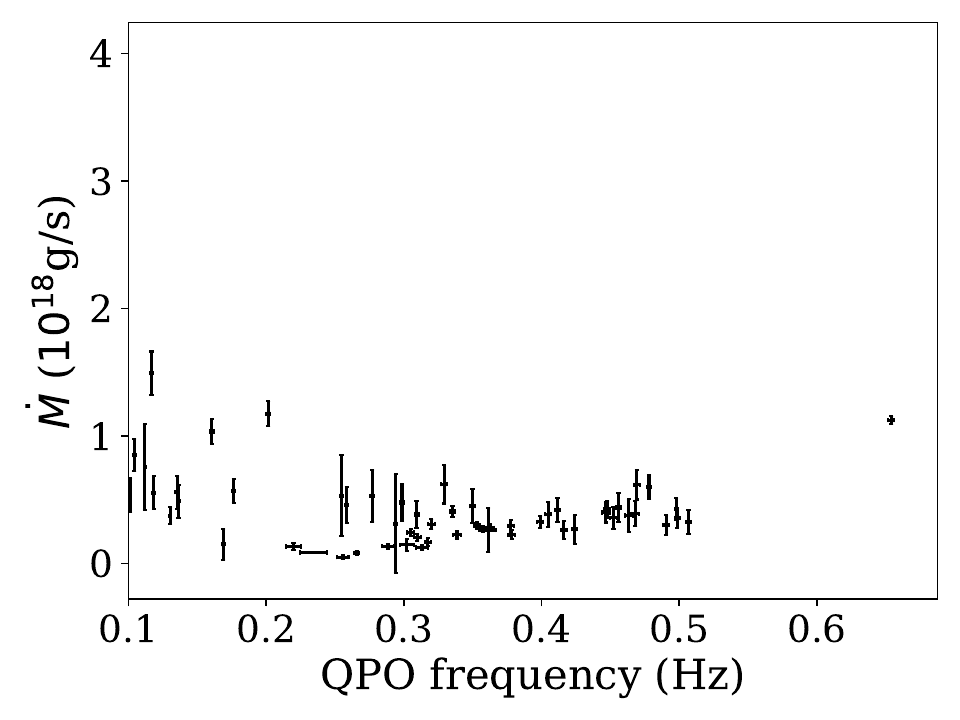}
        \includegraphics[width=0.32\textwidth]{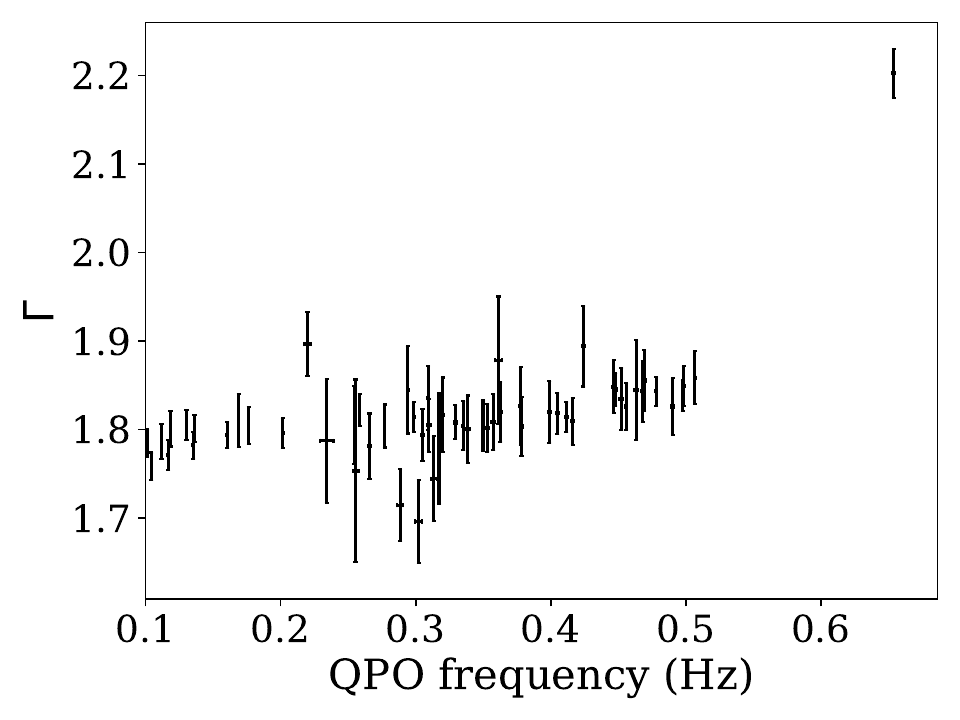}
    \caption{LFQPO frequency vs inner disk radius (left panel), LFQPO frequency vs mass accretion rate (middle panel), and 
    LFQPO frequency vs photon index (right panel). Only frequencies exceeding 0.1~Hz are reported.}
    \label{qpo_correlation}
\end{figure*}

Since how the QPO frequency changes with the evolution of spectral parameters can help us understand the origin of QPOs, we plot the relationship between the QPO frequency (obtained from \citealt{MaXiang}), the inner disk radius, mass accretion rate, and photon index derived from our fitting using Model 3 in Fig. \ref{qpo_correlation}. In the hard state, the mass accretion rate or the inner disk radius remains relatively stable, while the QPO frequency decreases, which is consistent with the findings of \cite{Buisson2019}, suggesting that the origin of LFQPOs in this scenario may not be directly linked to the instability of the mass accretion process. A positive correlation between the QPO frequency and photon index is found (the correlation coefficient is  about 0.65), indicating that the origin of QPOs may be related to changes in the corona. This correlation has also been discovered in studies of other sources \citep{Vignarca2003,zuobin_QPO}.

\subsection{Spin measurement}

In the soft state, for an accretion luminosity between $\sim 5$\% and $\sim 30$\% of the Eddington limit of the source, the disk is expected to extend to the ISCO \citep{2010ApJ...718L.117S,2011MNRAS.414.1183K}, therefore we can roughly estimate the BH spin through the inner disk radius in the soft state. In this section, we try to measure the spin value. We fit the spectra in the soft state with the models mentioned above by letting the spin to be a free parameter and freezing the inner disk radius to the ISCO. The best-fitting values and errors of the measured spin are shown in Table~\ref{spin_measurement}. 

\subsubsection{Reflection modeling}

In Model 2, we fit the reflection component with the relativistic reflection model \texttt{relxillCp} (in this model the thermal component is \texttt{diskbb}, which is non-relativistic, therefore there's no spin parameter in the thermal component) and measure the spin to be around 0.98 (see column 2 in Table~\ref{spin_measurement}). \cite{Draghis_spin} found that the spin of this source was $a_{*}=0.988^{+0.006}_{-0.028}$ with reflection modeling using NuSTAR data in both the hard and the soft state, consistent with our result.

\subsubsection{Continuum fitting}

In Models 3 and 3$^\star$, we measure the spin through the continuum fitting of the thermal component using the relativistic thermal component \texttt{nkbb} and reflection component \texttt{relxillCp} (the spin parameter in these two components are linked together). We find that different settings of parameters give completely different results. 

In Model~3, when the inclination angle is set to be $63^{\circ}$ \citep[which is the jet inclination measured in][]{Atri_2020}, the measured spin is about 0.3 (see column 4 in Table~\ref{spin_measurement}). However, in Model~3$^\star$, when the inclination angle is set to $32^{\circ}$ \citep[which is the disk inclination measured in][]{low_inclination}, the spin is about 0.94 (see column 3 in Table~\ref{spin_measurement}), which is more consistent with the result given by reflection modeling. 

\cite{Zhao_spin} and \cite{Guan_spin} measured the spin value to be $a_{*}=0.14\pm0.09$ and $a_{*}=0.2^{+0.2}_{-0.3}$ respectively by fitting the relativistic thermal component with \texttt{kerrbb2}, with the inclination angle set to be $63^{\circ}$. Fig. 19 in \cite{Draghis_spin} shows almost all current measurements of BH spins in X-ray binaries, obtained through continuum fitting or analysis of reflection features. MAXI J1820+070 is one of the sources in which continuum-fitting method and X-ray reflection spectroscopy provide inconsistent results.

\subsubsection{Relativistic dynamic frequency}

\begin{figure*}
    \centering
    \includegraphics[width=0.5\textwidth]{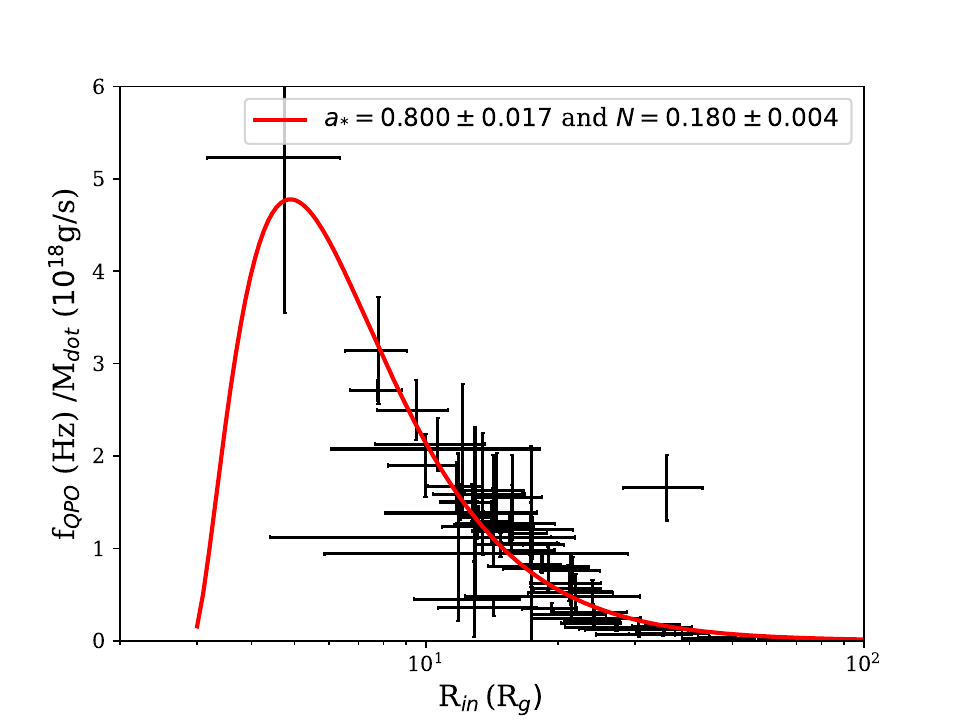}
    \caption{Plot $f_{\mathrm{QPO}}/\dot{M}$ vs inner disk radius $R_{\rm in}$. We fit the data with Equation~(3) in \cite{GRS1915}. The red solid curve is the best-fit with $a_{*}=0.800\pm0.017$ and $N=0.180\pm0.004$.}
    \label{qpo_fit}
\end{figure*}

Moreover, we follow the approach of \cite{GRS1915} and fit the relation between $f_{\mathrm{QPO}}/\dot{M}$ and $R_\mathrm{in}$ with the equation
\begin{equation}
    \begin{aligned}
         \frac{f_{\text {dyn}}}{\dot{M}_{18}}=&N\,8979 \mathrm{~Hz}\,\left(R_\mathrm{in}/R_\mathrm{g}\right)^{-2.5}\left(M / 12.4 M_{\odot}\right)^{-2}\\
        &\times A^1 B^{-2} D^{-0.5} E^{-0.5} L,   
    \end{aligned}
\end{equation}
where $R_{\mathrm{g}}$ is gravitational radius, $\dot{M}_{18}$ is the accretion rate in units of $10^{18}$~g/s, and the relativistic terms $A,B,D,E,L$ are functions of the radius and spin, and can be found in \cite{Novikov}. In \cite{GRS1915} it was reported that the correlation between $f_{\mathrm{QPO}}/\dot{M}$ and $R_{\rm in}$ could be fit well with this formula describing the relativistic dynamic frequency of a truncated accretion disk. Therefore, they identified the QPO frequency of GRS 1915+105 as the relativistic dynamic frequency of a truncated accretion disk. In our case, the fitting gives the spin $a_{*}=0.800\pm0.017$ and the normalization factor $N=0.180\pm0.004$ (see Fig.~\ref{qpo_fit}), consistent with $a_{*}=0.799^{+0.016}_{-0.015}$ measured by \cite{Bhargava_2021}. They fit the NICER data with relativistic precession model. This spin value is different from the measurement from both the fitting of the relativistic reflection \citep{Draghis_spin} and the continuum fitting of the thermal component \citep{Guan_spin,Zhao_spin}. However, spin measurements obtained from QPO frequencies should be taken
 with great caution, as we do not know yet the actual mechanism responsible for the observed QPOs and different QPO models normally provide different spin measurements.

\begin{table*}
\centering
    \begin{tabular}{cccc}
        Date & Model 2 ($i=63^{\circ}$) & Model 3$^\star$ ($i=32^{\circ}$) & Model 3 ($i=63^{\circ}$) \\
        \hline
        20180706 & $0.991^{+0.002}_{-0.003}$ & $0.977^{+0.003}_{-0.002}$ & $0.539^{+0.009}_{-0.013}$ \\
        20180707 & $0.99^{+0.002}_{-0.04}$ & $0.983^{+0.005}_{-0.003}$ & $0.48^{+0.02}_{-0.02}$ \\
        20180708 & $0.986^{+0.006}_{-0.002}$ & $0.953^{+0.006}_{-0.003}$ & $0.397^{+0.008}_{-0.008}$ \\
        20180709 & $0.989^{+0.002}_{-0.006}$ & $0.951^{+0.006}_{-0.003}$ & $0.339^{+0.018}_{-0.006}$ \\
        20180710 & $0.986^{+0.004}_{-0.002}$ & $0.936^{+0.006}_{-0.003}$ & $0.321^{+0.007}_{-0.001}$ \\
        20180711 & $0.988^{+0.005}_{-0.002}$ & $0.935^{+0.006}_{-0.003}$ & $0.308^{+0.011}_{-0.005}$ \\
        20180712 & $0.985^{+0.006}_{-0.005}$ & $0.947^{+0.007}_{-0.004}$ & $0.323^{+0.007}_{-0.008}$ \\
        20180713 & $0.989^{+0.002}_{-0.004}$ & $0.941^{+0.004}_{-0.002}$ & $0.323^{+0.004}_{-0.019}$ \\
        20180715 & $0.990^{+0.002}_{-0.003}$ & $0.938^{+0.003}_{-0.002}$ & $0.267^{+0.012}_{-0.006}$ \\
        20180717 & $0.986^{+0.005}_{-0.003}$ & $0.937^{+0.005}_{-0.002}$ & $0.301^{+0.005}_{-0.008}$ \\
        20180718 & $0.988^{+0.003}_{-0.003}$ & $0.939^{+0.006}_{-0.004}$ & $0.294^{+0.007}_{-0.013}$ \\
        20180719 & $0.986^{+0.002}_{-0.004}$ & $0.940^{+0.005}_{-0.002}$ & $0.317^{+0.008}_{-0.010}$ \\
        20180721 & $0.985^{+0.005}_{-0.003}$ & $0.938^{+0.005}_{-0.003}$ & $0.314^{+0.003}_{-0.003}$ \\
        20180722 & $0.987^{+0.011}_{-0.005}$ & $0.936^{+0.016}_{-0.006}$ & $0.26^{+0.04}_{-0.010}$ \\
        20180723 & $0.988^{+0.002}_{-0.002}$ & $0.951^{+0.003}_{-0.003}$ & $0.275^{+0.004}_{-0.010}$ \\
        20180725 & $0.991^{+0.005}_{-0.006}$ & $0.929^{+0.007}_{-0.005}$ & $0.247^{+0.014}_{-0.013}$ \\
        20180726 & $0.991^{+0.003}_{-0.006}$ & $0.929^{+0.006}_{-0.003}$ & $0.244^{+0.016}_{-0.004}$ \\
        20180727 & $0.987^{+0.004}_{-0.004}$ & $0.929^{+0.005}_{-0.003}$ & $0.299^{+0.003}_{-0.006}$ \\
        20180728 & $0.991^{+0.006}_{-0.003}$ & $0.935^{+0.004}_{-0.003}$ & $0.297^{+0.018}_{-0.009}$ \\
        20180731 & $0.988^{+0.008}_{-0.001}$ & $0.947^{+0.009}_{-0.008}$ & $0.317^{+0.007}_{-0.006}$ \\
        20180801 & $0.989^{+0.003}_{-0.004}$ & $0.941^{+0.019}_{-0.004}$ & $0.291^{+0.011}_{-0.013}$ \\
        20180802 & $0.985^{+0.003}_{-0.002}$ & $0.941^{+0.005}_{-0.004}$ & $0.301^{+0.003}_{-0.005}$ \\
        20180804 & $0.983^{+0.003}_{-0.002}$ & $0.947^{+0.004}_{-0.002}$ & $0.292^{+0.009}_{-0.012}$ \\
        20180805 & $0.991^{+0.006}_{-0.002}$ & $0.928^{+0.010}_{-0.005}$ & $0.29^{+0.03}_{-0.013}$ \\
        20180806 & $0.985^{+0.002}_{-0.003}$ & $0.942^{+0.007}_{-0.004}$ & $0.296^{+0.010}_{-0.017}$ \\
        20180808 & $0.989^{+0.002}_{-0.003}$ & $0.946^{+0.005}_{-0.003}$ & $0.31^{+0.02}_{-0.014}$ \\
        20180809 & $0.989^{+0.002}_{-0.003}$ & $0.94^{+0.02}_{-0.003}$ & $0.324^{+0.014}_{-0.010}$ \\
        20180811 & $0.984^{+0.005}_{-0.009}$ & $0.952^{+0.005}_{-0.002}$ & $0.316^{+0.011}_{-0.015}$ \\
        20180813 & $0.984^{+0.003}_{-0.005}$ & $0.952^{+0.005}_{-0.003}$ & $0.31^{+0.03}_{-0.02}$ \\
        20180815 & $0.985^{+0.005}_{-0.002}$ & $0.933^{+0.005}_{-0.003}$ & $0.27^{+0.01}_{-0.03}$ \\
        20180817 & $0.984^{+0.006}_{-0.005}$ & $0.945^{+0.007}_{-0.005}$ & $0.27^{+0.08}_{-0.03}$ \\
        20180819 & $0.986^{+0.003}_{-0.003}$ & $0.925^{+0.012}_{-0.006}$ & $0.27^{+0.14}_{-0.01}$ \\
        20180822 & $0.979^{+0.002}_{-0.004}$ & $0.944^{+0.007}_{-0.005}$ & $0.296^{+0.017}_{-0.007}$ \\        
        20180826 & $0.981^{+0.017}_{-0.007}$ & $0.933^{+0.011}_{-0.005}$ & $0.29^{+0.03}_{-0.02}$ \\
        20180829 & $0.979^{+0.011}_{-0.007}$ & $0.937^{+0.012}_{-0.008}$ & $0.30^{+0.008}_{-0.03}$ \\
        20180831 & $0.989^{+0.008}_{-0.009}$ & $0.931^{+0.008}_{-0.005}$ & $0.20^{+0.04}_{-0.01}$ \\
    \end{tabular}
    \caption{Spin measurements inferred with different models from the observations in the soft state. The dates of the spectra are in column~1, the spin measurements from the relativistic reflection components in Model~2 (where the inclination angle is $i=63^{\circ}$ and the disk electron density is $n_{e}=10^{15}$~cm$^{-3}$) are in column~2, the spin measurements from the relativistic thermal component in Model~3$^\star$ (where $i=32^{\circ}$, the hardening factor is $h_d=1.6$, and $n_{e}=10^{20}$~cm$^{-3}$) are in column~3, and the spin measurements from the relativistic thermal component in Model 3 (where $i=63^{\circ}$, $h_d=1.7$, and $n_{e}=10^{15}$~cm$^{-3}$) are in column 4.}
    \label{spin_measurement}
\end{table*}

\section{Summary and conclusion}\label{conclusion}
In this study, we fit the spectra of the 2018 outburst of MAXI J1820+070 observed by \textit{Insight}-HXMT with different models. The main findings are reported in Table~\ref{model_comparison} and can be summarized as follows:

\begin{enumerate}
    \item {Models with different thermal components give different fitting results of the inner disk radius and spin values in the soft state. Model 2, adopting the non-relativistic thermal model \texttt{diskbb}, shows that $R_{\mathrm{in}}$ is near the ISCO and $a_*$ is about 0.98. Model~3, adopting the relativistic thermal model \texttt{nkbb}, shows that $R_{\mathrm{in}}$ is about $10\,R_{\mathrm{ISCO}}$ and $a_*$ is about 0.3. However, if we set the inclination angle to be $32^{\circ}$ and change the hardening factor and disk electron density (Model~3$^\star$), the results given by the two models are more consistent. The spin inferred with Model~3$^\star$ is about 0.94.}
    
    \item {We study the evolution of the corona from the analysis of the scattering fraction and emissivity index $q$. All models agree that the scattering fraction is high in the hard and intermediate states and it is low in the soft state. In Model~2, the emissivity index $q$ is very high in the soft state, therefore the corona may be compact and close to the black hole. However, in Model~3 and Model~3$^\star$, the emissivity index $q$ remains almost stable in the whole outburst, indicating that the cause of the reduced scattering fraction in the soft state may be the changes in the corona electron density rather than in the corona size.} 
    
    \item We study the relationship among the inner disk radius, mass accretion rate, photon index and the LFQPO frequency. 
    We do not find any correlation between the LFQPO frequency and the inner disk radius and the mass accretion rate.
    We observe a positive correlation between the photon index and the QPO frequency. We employ the formula for the relativistic dynamic frequency of a truncated accretion disk and we obtain the black hole spin parameter $a_{*}=0.800\pm0.017$.
\end{enumerate}

\vspace{0.5cm}

{\bf Acknowledgments --}
This work was supported by the Natural Science Foundation of Shanghai, Grant No.~22ZR1403400, and the National Natural Science Foundation of China (NSFC), Grant No.~12250610185, 11973019, and 12261131497.

\bibliographystyle{aasjournal}
\bibliography{ref}

\end{document}